# RELATIONSHIP AMONG EFFICIENCY AND OUTPUT POWER OF HEAT ENERGY CONVERTERS

Dr. Alexander LUCHINSKIY

Johannes Gutenberg – University Mainz, Germany

Post: Gymnasialstr. 11, Bad Kreuznach, Germany, D-55543
E-mail: QHETS@AOL.com
Tel./Fax.: +49 671 / 35594

**Abstract**

Relationship among efficiency and output power of heat-electric energy converters as well as of any converters for transforming of heat energy into any other kind of energy is considered. It is shown, that the parameter efficiency does not determine univocally the output power of a converter. It is proposed to use an other parameter for determination of working ability of heat energy converters. It is shown, that high output power can not be achieved by any kind of Stirling-type converters in spite of their high efficiency.

It has been historically automatically believed that a principal theoretical constraint for output power of low-energy-density heat-electric converters is the temperature gradient between the cooling and heating parts of the converter, which is expressed by their efficiency equation: $\eta = (T_{input} - T_{output}) / T_{input}$, where $T_{input}$ and $T_{output}$ are the temperatures of external media at the heater and cooler of a heat-electrical converter respectively.

Nevertheless the above mentioned historical belief is, in fact, a formal misunderstanding. The efficiency equation does not determine the output power, but, rather, the theoretically possible maximum of the efficiency as in $E_{output}/E_{input}$. This ratio of output and input *energies* does not take into account time **t**, and, therefore, does not determine the output *power* $P_{output} = E_{output} / t$. For instance, if two heat-electrical converters are working with the same temperature gradient, then one of them can have a higher efficiency, but a lower output power, and vice versa. The key consideration here is how *quickly* the converter operates (which does not only depend on the temperature gradient). Quick conversion also helps to minimize losses through dissipation.

**Let us consider a heat – electric converter (or any other kind of converter, which one transforms a heat energy into an other kind of energy).**

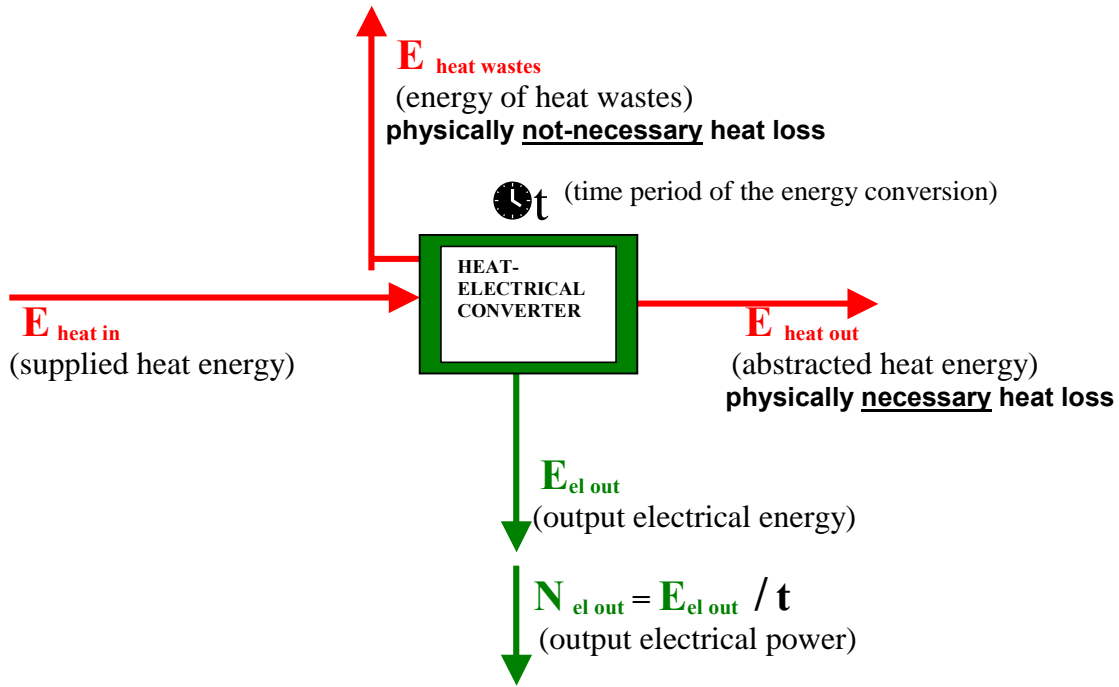

- $E_{heat\ wastes}$ (energy of heat wastes) **physically not-necessary heat loss**
- $t$ (time period of the energy conversion)
- **HEAT-ELECTRICAL CONVERTER**
- $E_{heat\ in}$ (supplied heat energy)
- $E_{heat\ out}$ (abstracted heat energy) **physically necessary heat loss**
- $E_{el\ out}$ (output electrical energy)
- $N_{el\ out} = E_{el\ out} / t$ (output electrical power)

Balance of the energy is: $\quad E_{el\ out} = E_{heat\ in} - E_{heat\ out} - E_{heat\ wastes}$

Output electrical power is: $\quad N_{el\ out} = E_{el\ out} / t$

$$N_{el\ out} = E_{heat\ in} - E_{heat\ out} - E_{heat\ wastes} / t$$

**therefore $N_{el\ out}$ can be maximised, if $E_{heat\ wastes}$ and $t$ will be minimised.**

---

## EXPRESSIONS FOR EFFICIENCY η AND $N_{el\ out}$ THROUGH η :

1) Efficiency η can be maximised, if the energy of heat wastes $E_{heat\ wastes}$ will be minimised:

$$\eta = E_{el\ out} / E_{heat\ in} = (E_{heat\ in} - E_{heat\ out} - E_{heat\ wastes}) / (E_{heat\ in})$$

2) Output electrical power can be maximised, if :

a) **Efficiency η will be maximised;**
b) **Time t of conversion will be minimised:**

$$N_{el\ out} = (E_{el\ out}) / t = (\eta \times E_{heat\ in}) / t = E_{heat\ in} \times (\eta / t)$$

**Conditions a) and b) for (η) and (t) are independent one from another. Theoretically and technically the variants are possible, when a converter with the lower efficiency η has the higher output electrical power $N_{el\ out}$ dependently of the value of time t:**

$$N_{el\ out} = E_{heat\ in} \times (\eta / t)$$

(here $E_{heat\ in}$ is the supplied to the converter heat energy; $E_{heat\ out}$ is the abstracted from the converter heat energy (physically necessary heat loss); $E_{heat\ wastes}$ is the heat energy of heat wastes (physically not-necessary heat loss); $E_{el\ out}$ is the output electrical energy of the converter; $t$ is the time; and $N_{el\ out}$ is the output electrical power of the converter).

---

**Additional general notes:**

It is necessary to note, that in the case of fossil energy sources the efficiency $\eta$ really plays a decided role, because it determines how much energy from the burned energy-carrier substance (fuel) will be utilised. But in the case of renewable heat energy sources finally it does not matter how much energy of a source will be loss, because there is no prise for the energy-carrier substance and there is no pollutions. Therefore efficiency $\eta$ plays a role as far as it influence on the output power. (If energy losses during the energy conversion will be reduced, the output power will increase). But as it was shown above, it is only one of parameters, but not the sole one, and not the main one.

Parameter $\eta$ (efficiency) does not determine a workingability, technically and commercially, of the heat-electrical converter adequately. It is necessary to propose an other parameter to describe it's effectiveness. It is proposed to use a parameter (coefficient)

$$l = N_{output} / \Delta T$$

that is a ratio of output power $N_{output}$ of a converter to the input temperature difference $\Delta T$ to describe the working characteristics of a renewable heat energy converters instead of the parameter $\eta$ (efficiency).

In fact both efficiency and velocity of energy conversion are taken into account by this parameter.

Therewith if this parameter (coefficient) is named for a converter, an output power of this converter by any of input temperature differences is determined.

**Conclusion 1:**
Parameter $\eta$ (efficiency) does not determine the output power of a converter univocally. Theoretically and technically the variants are possible, when a converter with the lower efficiency $\eta$ has the higher output electrical (or other non-heat) power $N_{el\ out}$, or when two converters with the same efficiency $\eta$ have different output electrical (or other non-heat) powers $N_{el\ out}$

**Conclusion 2 :**
Parameter $\eta$ (efficiency) does not determine adequately a workingability and utility of a heat-electrical converter (or any converter, which transforms a heat energy into other form of energy). It is proposed to use a parameter (coefficient) "$l = N_{output} / \Delta T$" (Ratio of output power of a converter to the input temperature difference) to describe the working characteristics of a renewable heat energy converters instead of the parameter $\eta$ (efficiency).

---

The above-described theoretical matter lead to one direct practical consequence: it is principally impossible to build any kind of Stirling – motor with high output power in spite of it's high efficiency. It is possible to essentially increase it's efficiency, but it will not lead to the creation of a practically applicable converter with high output power.

In fact, physically, the action of Stirling-motors is based on a slow absorption of heat energy by a slowly extending gas. The amount of energy absorbed and converted by the Stirling-motor in a unite of time is very small, and therefore it's output power is very low too. Efficiency is not an exhaustive parameter to describe effectiveness of the converter of energy, if there is no specification how much time spend this converter to convert a given amount of energy (or specification about it's output power).

Efficiency $\eta$ of the converter of energy is, as it is known, a ratio between it's input and output powers $N_{IN}$ and $N_{OUT}$ or the ratio between supplied and converted energies $E_{IN}$ and $E_{OUT}$.

Because power $N$ is energy $E$ in a unit of time $t$ $(N = E/t)$, efficiency is calculated as: $\eta = N_{IN} / N_{OUT} = (N_{IN} * t) / (N_{OUT} * t)$ where $N_{IN}$ and $E_{IN}$ are the input power and energy, and $N_{OUT}$ and $E_{OUT}$ are the output power and energy on the output of the converter. Time $t$ is both in a numerator and in a denominator, and is therefore cancelled. Thus, efficiency $\eta$ does not define time for which the given amount of energy was converted, but only how much energy was lost during this conversion. **That is the generators converting some energy for 1 minute and for 1 hour can have identical rated efficiency**. (Previous publication of this matter took place in [ISBN 3-8288-1255-4, "Erneuerbare Energiequellen: Komplex von Technischen Losungen ", A.Luchinski; Tectum - Wissenschaftsverlag, Marburg, 2002]; germ.).

**Conclusion:** Output power of Stirling-motors is very low, despite of high rated efficiency. This efficiency is normally calculated as a ratio between the converted output energy and the supplied input energy without taking into account time spent for this conversion.

It is necessary to note two significant factual circumstances:
1) In despite of the formally higher then in photovoltaical converters efficiency, the Stirling converters are not used as the house-roof sun converters for electricity house-supply instead of PV-panels;
2) Already now the efficiency of presently created Stirling motors much more higher then the efficiency of internal-combustion engines (according to some publications a level of 50% and even more efficiency for Stirling-motors is achieved). But nobody substitutes the internal-combustion engines with Stirling-motors, because the Stirling-motors have much lower output power in despite of higher efficiency.

The above presented consideration explains the reason of these facts and the matter of the presently taking place formal misunderstanding.

## Conclusions :

1. Efficiency-parameter does not determine the output power of a converter univocally. Theoretically and experimentally the variants are possible, when a converter with the lower efficiency has the higher output electrical (or other non-heat) power**,** or when two converters with the same efficiency have different output electrical (or other non-heat) powers**.**

2. Parameter $\eta$ (efficiency) does not determine adequately a workingability and utility of a heat- electrical converter (or any converter, which transforms a heat energy into other form of energy). It is proposed to use a parameter (coefficient) "$l = N_{output} / \Delta T$" (Ratio of output power of a converter to the input temperature difference) to describe the working characteristics of a renewable heat energy converters instead of the parameter $\eta$ (efficiency)**.**

3. As essential as one likes increasing of an efficiency of the Stirling-type converters through any kind of technical solutions can not lead to the essential increasing of their output power. It is principally impossible to build a Stirling-motor with high output power, also regardless of it's efficiency, as long as it's action is based on a slow absorption of heat energy by a slowly extending gas.

\* \* \*